\documentclass[aps,prl,reprint,showpacs]{revtex4-1}
\usepackage{graphicx}
\usepackage{dcolumn}
\usepackage{bm}

\begin{document}

\title{The light filament as a new nonlinear polarization state}

\author{Lubomir M. Kovachev}
\affiliation{Institute of Electronics, Bulgarian Academy of Sciences,\\
Tzarigradcko shossee 72,1784 Sofia, Bulgaria}
\email[]{lubomirkovach@yahoo.com}
\date{\today}

\begin{abstract}
We present an analytical approach to the theory of nonlinear
propagation in gases of femtosecond optical pulses with broad-band
spectrum . The vector character of the nonlinear third-order
polarization of the electrical field in air is investigated in
details. A new polarization state is presented by using left-hand
and right-hand circular components of the electrical field . The
corresponding system of vector amplitude equations is derived in the
rotating basis. We found that this system  of nonlinear equations
has $3D+1$ vector soliton solutions with Lorentz shape. The solution
presents a relatively stable propagation and rotation with GHz
frequency of the vector of the electrical field in a plane
orthogonal to the direction of propagation. The evolution of the
intensity profile demonstrates a weak self-compression and a week
spherical wave in the first milliseconds of  propagation.
\end{abstract}

\maketitle

\section{Introduction}

When a femtosecond  laser pulse with power above the critical for
self-focusing propagates in air, a number of new physical effects
are observed, such as long-range self-channeling \cite{MOU,WO},
coherent and incoherent GHz and THz emission \cite{TZ, DAMYZ},
asymmetric pulse shaping, super-broad spectra \cite{HAU, CHIN1},
polarization instability of linearly polarized pulse \cite{MOL},
polarization rotation \cite{KOZA}, self-compression \cite{KOPR} and
others. A remarkable effect is also that in lidar experiments the
light filaments propagate over distances up to 9-10 kilometers in
vertical direction, preserving their spectrum and shape \cite{WO}.
In a typical experiment in the near zone (up to $1-3$ $m$ from the
source), when the initial pulse intensity exceeds $I>10^{12}
W/cm^2$, self-focusing and self-compressing start, which makes the
$k_z$ spectrum broad band and asymmetric $\triangle k_z\approx k_0$.
The process increases the core intensity up to $10^{14} W/cm^2$,
where a short non-homogenous plasma column in the nonlinear focus is
observed. Usually the standard model describing the propagation in
the near zone is a scalar spatio-temporal nonlinear paraxial
equation including in addition terms with plasma ionization, higher
order Kerr terms, multiphoton ionization and others \cite{KAN,
KASP2}. The basic model works properly in the near zone because of
the fact that the paraxial approximation is valid for pulses with
narrow-band spectrum $\triangle k_z << k_0$. At distances longer
than 10-20 meters from the laser source, where the stable filament
is formed, plasma generation and higher-order Kerr terms are also
included as necessary for the balance between the self-focussing and
plasma defocussing and for obtaining long range self-channeling in
gases. However, the above explanation of the filamentation is
difficult to be applied  at such distances. As reviewed in
\cite{MECH,MECH2, KASP2, KILO,SHIM,KAN1,KAN2}, the plasma density at
long distances from the source is too small to prevent
self-focusing. There are basically three main characteristics which
remain unchanged at these distances - the broad-band spectrum, the
coherent GHz generation and the width of the core, while the
intensity in a stable filament drops to a value of $10^{12} W/cm^2$.
The higher-order Kerr terms for pulses with intensities of the order
of $I\sim 10^{12} W/cm^2$  are also too small to prevent
self-focussing. The experiments, where observation of long-range
self-channeling without ionization was realized \cite{MECH,
MECH2,KILO}, show the need to change the role of the plasma
defocusing at such distances with another effect. In addition, there
are difficulties with the physical interpretation of the coherent
GHz radiation as a result of plasma generation. The light pulse near
the nonlinear focus emits incoherent and \emph{non-homogenous}
plasma, while the coherent GHz radiation requires \emph{homogenous}
plasma with fixed electron density of the order of $10^{15}$
$cm^{-3}$. Only homogenous plasma can generate coherent GHz
emission, but such kind of plasma is absent in the process of
filamentation. In the real experiments with propagation of a single
filament at distances more than $20-30$ $m$ from the source in air,
the following basic characteristics  are found:

1. Broad-band spectrum ($\Delta k_z \sim k_0$).

2. Intensity of the order of $I \simeq 10^{11-12}$ $W/cm^2$.

3. Absence of plasma at long distances.

4. Asymmetric relatively stable (Lorentz) spectral and longitudinal
shapes.

5. Coherent GHz generation.

Recently in \cite{KOVBOOK} we developed a scalar ionization-free
non-paraxial nonlinear model, which gives the above characteristics
of the stable filament. The analytical and the numerical results
describe correctly the linear and nonlinear evolution of narrow-band
and broad-band laser pulses. In addition it was found that the
equation has exact Lorentz-type soliton solutions in approximation
of neglecting the GHz oscillation. Still, this theory cannot resolve
some difficulties. The main problems are:

1. Peak instability of the soliton solution under small initial
perturbations.

2.  The soliton solution is obtained after neglecting the GHz
oscillation.

3. The soliton solution has one free parameter.

4. There are problems with the conservation law of the nonlinear
operator when we use the GHz oscillation.

To solve the above problems in this paper, we propose  a nonlinear
vector generalization to the model.

\section {Nonlinear Polarization}

The self-action process broadens the pulse spectrum - starting from
a narrow-band pulse, the stable filament becomes broad-band far from
the source. In recent papers  \cite{KOL2,LMK1} it is shown that the
evolution of broad-band pulses like filaments can be described
correctly by useing the generalized nonlinear polarization

\begin{eqnarray}
\label{NLTH} \vec{P}^{nl} = n_2 \left( \vec{E} \cdot \vec{E}
\right)\vec{E},
\end{eqnarray}
which includes additional processes associated with third harmonic
generation (THG). A more precise analysis presented in \cite{KOVGEO}
demonstrates that the polarization of the kind (\ref{NLTH}) is not
applicable to a scalar model, because the corresponding Manley-Rowe
(MR) conservation laws are not satisfied. That is why we investigate
two-component electrical vector field. The generalized nonlinear
polarization (\ref{NLTH}) is quite simple in terms of left-hand and
right-hand circular components. Let us now present the electric
vector field of a pulse as  a linear decomposition of left- and
right-hand circular complex components.

\begin{eqnarray}
\label{ELF1} \vec{E}(x,y,z,t) = E_+(x,y,z,t)\vec{\sigma}_{+} +
E_-(x,y,z,t)\vec{\sigma}_{-}
\end{eqnarray}
where  the circular- polarization unit vectors are
\begin{eqnarray}
\label{CIRKKOMP} \sigma_{\pm}=(\vec{x}\pm i\vec{y})/\sqrt{2}.
\end{eqnarray}
If we now represent $\vec{P}^{nl}$ in terms of its circular
components as

\begin{eqnarray}
\label{PNL} \vec{P} = P_+\vec{\sigma}_{+} + P_-\vec{\sigma}_{-},
\end{eqnarray}
we find that the components  are given by

\begin{eqnarray}
\label{NLCIRK} P_+ =
2n_2\left(E_+^2E_-\right)\\
P_- = 2n_2 \left(E_-^2E_+\right).
\end{eqnarray}

\section{Basic System of Equations}
The decomposition (\ref{ELF1}) allows us to rewrite the nonlinear
vector wave equation in the following system of equations.

\begin{eqnarray}
\label{BBWAVE} \Delta E_{\pm} -\frac{1}{c^2}\frac{\partial^2
E_{\pm}}{\partial t^2}=\frac{4\pi}{c^2}\frac{\partial^2}{\partial
t^2}\Big\{\int_{0}^{\infty}R^{(1)}(\tau)E_{\pm}(t-\tau)d\tau\nonumber\\
\\
+\int_{0}^{\infty}\int_{0}^{\infty}\int_{0}^{\infty}R^{(3)}(\tau,\tau,\tau)\left[E_{\pm}(t-\tau)\right]^2E_{\mp}(t-\tau)d\tau^3
\Big\}\nonumber
\end{eqnarray}
To obtain amplitude equations, we use in mind also the causality
principles (no negative time $\tau$ ) to the response functions and
their Fourier presentations $\chi^{(1)}(\omega)$   and
$\chi^{(3)}(\omega, \omega, \omega)$

\begin{eqnarray}
\label{omega} \chi^{(1)}(\omega)
=\int_{0}^{\infty}R^{(1)}(\tau)\exp(i\omega\tau)d\tau \nonumber\\
\\
\chi^{(3)}(\omega, \omega, \omega)
=\int_{0}^{\infty}\int_{0}^{\infty}\int_{0}^{\infty}R^{(3)}(\tau,
\tau, \tau)\exp(3i\omega\tau)d\tau^3.\nonumber
\end{eqnarray}
Reduction of the integrals (\ref{omega}) from $0$ to infinity is
equal to $\cos$Fourier transforms with the properties

\begin{eqnarray}
\label{omega1}
\chi^{(1)}(\omega) =\chi^{(1)}(-\omega)\nonumber\\
\\
\chi^{(3)}(\omega, \omega, \omega)= \chi^{(3)}(-\omega, \omega,
\omega)= \chi^{(3)}(-\omega, -\omega, \omega)...\nonumber
\end{eqnarray}

The key point in the following transformations is the fact that the
circular components of the electrical field, as well as the
amplitude functions, are orthogonal in the complex plane. That is
why their Fourier presentations are also written in orthogonal basis

\begin{equation}
\label{omega2}
E_{+}(t-\tau)=\int_{-\infty}^{\infty}\hat{E_+}(\omega)\exp(-i\omega(t-\tau)d\omega
\end{equation}
\begin{equation}
\label{omega3}
E_{-}(t-\tau)=\int_{-\infty}^{\infty}\hat{E_-}(\omega)\exp(i\omega(t-\tau)d\omega.
\end{equation}
Substituting (\ref{omega2}) - (\ref{omega3}) into the right-hand
side and also to the last term of left-hand side in (\ref{BBWAVE}),
and using the spectral properties (\ref{omega1}) of the response
functions (\ref{omega}), the wave system can be written as
\begin{eqnarray}
\label{WWAVE1} \Delta
E_{+}=-\int_{-\infty}^{\infty}k^2(\omega)\hat{E}_{+}(\omega)\exp(-i\omega
t)d\omega\nonumber\\
\\
-\int_{-\infty}^{\infty}\int_{-\infty}^{\infty}\int_{-\infty}^{\infty}k_{nl}^2(\omega)\left[\hat{E}_{+}\right]^2\hat{E}_{-}\exp(-i\omega
t)d\omega^3\nonumber
\end{eqnarray}

\begin{eqnarray}
\label{WWAVE2} \Delta
E_{-}=-\int_{-\infty}^{\infty}k^2(\omega)\hat{E}_{-}(\omega)\exp(i\omega
t)d\omega\nonumber\\
\\
-\int_{-\infty}^{\infty}\int_{-\infty}^{\infty}\int_{-\infty}^{\infty}k_{nl}^2(\omega)\left[\hat{E}_{-}\right]^2\hat{E}_{+}\exp(i\omega
t)d\omega^3\nonumber,
\end{eqnarray}
where $k^2(\omega)=\omega^2\varepsilon(\omega)/c^2$;
$k^2_{nl}(\omega)=\omega^2\chi^{(3)}(\omega, \omega,
\omega)/c^2=k^2(\omega)\hat{n}_2(\omega)$ are the square of the
linear and nonlinear wave vectors and

\begin{eqnarray}
\label{epsn2}
\varepsilon(\omega)=1+4\pi\chi^{(1)}(\omega)\\
n_2(\omega)=4\pi\chi^{(3)}(\omega,\omega,\omega)/\varepsilon(\omega).
\end{eqnarray}
Lets us now introduce complex amplitude functions using the
substitutions

\begin{eqnarray}
\label{Amp1}
E_{+}(r,t)=A_{+}(r,t)\exp\left[i\left(k_0z-\omega_0t\right)\right]
\end{eqnarray}

\begin{eqnarray}
\label{Amp2}
E_{-}(r,t)=A_{-}(r,t)\exp\left[-i\left(k_0z-\omega_0t\right)\right].
\end{eqnarray}
where $r=(x,y,z)$, $\omega_0$ is the carrier frequency and $k_0$ is
the linear part of the wavevector ar the carrier frequency. Applying
the translation theorem to the Fourier presentations of the
electrical field components we have

\begin{eqnarray}
\label{Ampw1} \hat{E}_{+}(r,\omega)=\exp
(ik_0z)\hat{A}_{+}(r,\omega-\omega_0)
\end{eqnarray}

\begin{eqnarray}
\label{Ampw2} \hat{E}_{-}(r,\omega)=\exp
(-ik_0z)\hat{A}_{-}(r,\omega-\omega_0).
\end{eqnarray}
Substituting (\ref{Amp1})-(\ref{Ampw2}) into Eqs. (\ref{WWAVE2}) we
obtain

\begin{eqnarray}
\label{AM1} \Delta A_{+}+2ik_0\frac{\partial A_{+}}{\partial
z}-k_0^2A_{+}=\nonumber\\
-\int_{-\infty}^{\infty}k^2(\omega)\hat{A}_{+}\exp\left[-i(\omega-\omega_0)t\right]
t)d\omega\nonumber\\
\\
-\int\int\int_{-\infty}^{\infty}k_{nl}^2(\omega)
\left[\hat{A}_{+}\right]^2\hat{A}_{-}\exp\left[-i(\omega-\omega_0)t\right]d\omega^3\nonumber
\end{eqnarray}

\begin{eqnarray}
\label{AM2} \Delta A_{-}-2ik_0\frac{\partial A_{-}}{\partial
z}-k_0^2A_{+}=\nonumber\\
-\int_{-\infty}^{\infty}k^2(\omega)\hat{A}_{-}\exp\left[i(\omega-\omega_0)t\right]
t)d\omega\nonumber\\
\\
-\int\int\int_{-\infty}^{\infty}k_{nl}^2(\omega)
\left[\hat{A}_{-}\right]^2\hat{A}_{+}\exp\left[i(\omega-\omega_0)t\right]d\omega^3\nonumber
\end{eqnarray}
We note that all functions in the integrals on the right side of
Eqs. (\ref{AM1}) and (\ref{AM2}) depend form on the  frequency
difference $(\omega-\omega_0)$, except for the linear $k^2(\omega)$
and nonlinear $k^2_{nl}(\omega)$ wave vectors. Then we expand
$k^2(\omega)$ and $k^2_{nl}(\omega)$ as a power series of the same
difference $(\omega-\omega_0)$

\begin{eqnarray}
\label{K2}
k^2(\omega)=k_0^2+2\frac{k_0}{v_{gr}}\left(\omega-\omega_0\right)\nonumber\\
+\left(\frac{1}{v_{gr}^2}+k_0k''\right)(\omega-\omega_0)+...
\end{eqnarray}

\begin{eqnarray}
\label{K22}
k^2_{nl}(\omega)=k^2(\omega)\hat{n}_2(\omega)=k_0^2n_2|_{\omega=\omega_0}\nonumber\\
+\left[\frac{2k_0}{v_{gr}}n_2(\omega_0)+k_0^2\frac{\partial
n_2}{\partial\omega}|_{\omega=\omega_0}\right]\left(\omega-\omega_0\right)+...,
\end{eqnarray}
where  $v_{gr}$ is the group  velocity,  and $k''$ is the group
velocity dispersion. After replacing $k^2(\omega)$ and
$k^2_{nl}(\omega)$ with their series (\ref{K2})-(\ref{K22}), all
functions in the integrals from the right side of Eqs. (\ref{AM1})
and (\ref{AM2}) depend on the frequency difference and we can
integrate over all values of $(\omega-\omega_0)$. Thus, we obtain
the following system of amplitude equations in circular basis (up to
second order of the group velocity dispersion and zero order of the
nonlinear dispersion).
\begin{eqnarray}
\label{SYSCIRK1} 2ik_0 \left( \frac{\partial A_+}{\partial
z}+\frac{1}{v_{gr}}\frac{\partial A_+}{\partial t}\right)=
 \Delta A_+ - \frac{1+\beta}{v_{gr}^2}\frac{\partial^2 A_+}{\partial t^2}\nonumber\\
+ 2k^2_0 n_2A^2_+A_-\nonumber\\
\\
-2ik_0 \left( \frac{\partial A_-}{\partial
z}+\frac{1}{v_{gr}}\frac{\partial A_-}{\partial t}\right)=
 \Delta A_- - \frac{1+\beta}{v_{gr}^2}\frac{\partial^2 A_-}{\partial t^2}\nonumber\\
+2 k^2_0 n_2A^2_-A_+,\nonumber
\end{eqnarray}
where  $\Delta$ is $3D$ - $(x,y,z)$ Laplace operator,  and $\beta$
is a number, connected with the dispersion characteristics of the
medium ($\beta = k_0v_{gr}^2k''$). We note here that in gases the
dispersion is weak and the series (\ref{K2}) are strongly converged
up to single  cycle regime (broad-band pulses). Thus, the
non-paraxial system of equations (\ref{SYSCIRK1}) describe correctly
the evolution of laser pulses in gases up to a single cycle regime.
It is important to  mentoin that from Eqs. (\ref{SYSCIRK1}) paraxial
spatio-temporal approximation can be derived  for narrow-band laser
pulses \cite{KOVBOOK} only. The filamentation experiments
demonstrate quite different pulse evolution: the initial laser pulse
$\left( t_0\geq 50 fs \right) $ possesses a relatively narrow-band
spectrum $\left(\Delta k_z \ll k_0 \right)$ and during the process
the initial self-focusing and self-compression  the spectrum
broadens significantly. The broad-band spectrum $\left(\Delta k_z
\sim k_0\right)$ is one of the basic characteristics of the stable
filament. That why we do not reduce more Eqs. (\ref{SYSCIRK1}) and
try to solve them for the case when the pulse has a large spectrum.

Another standard restriction in the filamentation theory is the use
of one-component scalar approximation of the electrical field
$\vec{E}$. This approximation, though,  is in contradiction with
recent experimental results, where rotation of the polarization
vector is observed \cite{KOZA}. For this reason in the present paper
we use non-paraxial vector model in circular basis (\ref{SYSCIRK1}),
in which the nonlinear effects are described by the nonlinear
polarization components (\ref{NLCIRK}). The dispersion number in air
is very small: $\beta = k_0v_{gr}^2k''\simeq 2.1\times10^{-5}$, so
we can solve Eqs. (\ref{SYSCIRK1}) in approximation up to  the first
order of dispersion. Additionally, we will use the normalized
amplitude functions $A_{\pm}=A_0A_{\pm}$ to rewrite the system
(\ref{K2}) in the form

\begin{eqnarray}
\label{SYSCIRK} 2ik_0 \left( \frac{\partial A_+}{\partial
z}+\frac{1}{v_{gr}}\frac{\partial A_+}{\partial t}\right)=
 \Delta A_+ - \frac{1}{v_{gr}^2}\frac{\partial^2 A_+}{\partial t^2}\nonumber\\
+ \gamma A^2_+A_-\nonumber\\
\\
-2ik_0 \left( \frac{\partial A_-}{\partial
z}+\frac{1}{v_{gr}}\frac{\partial A_-}{\partial t}\right)=
 \Delta A_- - \frac{1}{v_{gr}^2}\frac{\partial^2 A_-}{\partial t^2}\nonumber\\
+ \gamma A^2_-A_+,\nonumber
\end{eqnarray}
where $\gamma=2n_2k_0^2A_0^2$ is a nonlinear coefficient.

\begin{figure*}
\centerline{\includegraphics[width=140mm,height=85mm]{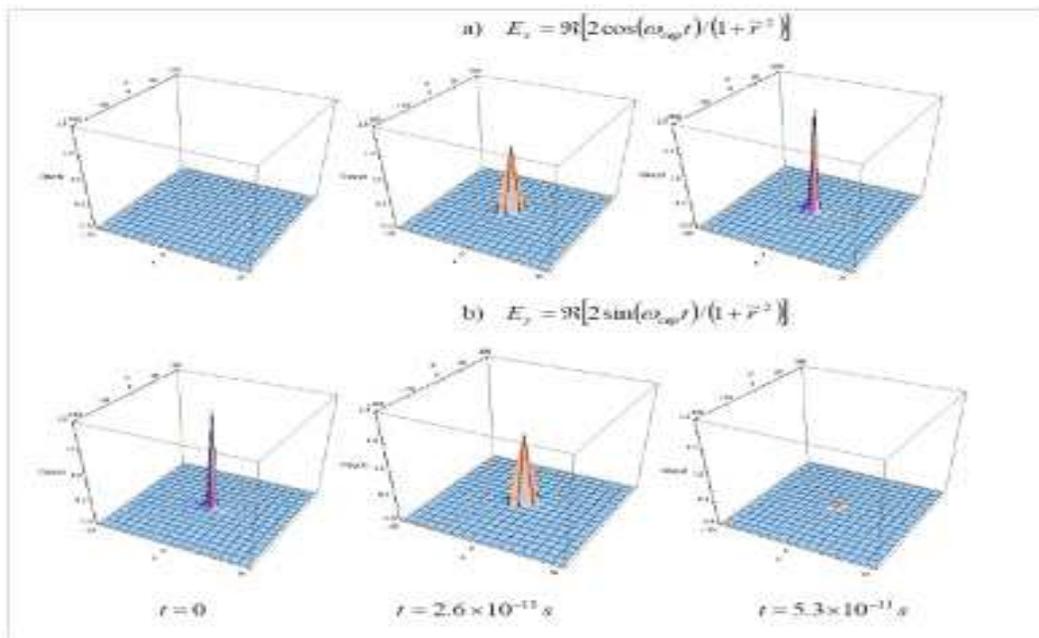}}
\caption{ Evolution of the  profiles of the electrical field
components $\Re (E_x)$-Fig. 1a and $\Re(E_y)$-Fig. 1b  of the
solution (\ref{REfield}). The periodical exchange of energy between
components, due to nonlinear mechanism, leads to rotation of the
electrical field vector in a plane orthogonal to the direction of
propagation.}
\end{figure*}

\section{Vector solution and vector rotation. The filament as a weak Rogue wave}

The nonlinear system of equations (\ref{SYSCIRK}) has exact solitary
vector solution when $\gamma=2$ and the spectral width of the pulses
reaches the value $\Delta k_z\approx k_0$

\begin{eqnarray}
\label{soliton1}
A_+(x,y,z,t)=
\frac{2}{1+\tilde{r}^2}\exp\left[i\Delta k_z\left(z-v_{gr}t\right)\right]\nonumber\\
\\
A_-(x,y,z,t)=\frac{2}{1+\tilde{r}^2}\exp\left[-i\Delta
k_z\left(z-v_{gr}t\right)\right],\nonumber
\end{eqnarray}
where
$\tilde{r}=\sqrt{x^2+y^2+(z-ik_{cef})^2-v_{gr}^2(t-ik_{cef}/v_{gr})^2}$
and $k_{cef}$ is determinated below. The solution of the
corresponding vector electrical field can be written after
multiplying the amplitude functions (\ref{soliton1}) by the main
phases (\ref{Amp1})-(\ref{Amp2})

\begin{eqnarray}
\label{Efield}
E_+(x,y,z,t)=\frac{2}{1+\tilde{r}^2}
\exp\left[i\Delta k_z\left(v_{ph}-v_{gr}\right)t\right]\nonumber\\
\\
E_-(x,y,z,t)=\frac{2}{1+\tilde{r}^2}\exp\left[-i\Delta
k_z\left(v_{ph}-v_{gr}\right)t\right].\nonumber
\end{eqnarray}
Let us turn from the left-hand and right-hand circular components
(\ref{CIRKKOMP}) to the standard Cartesian coordinates

\begin{eqnarray}
\label{CART} E_x = (E_+ +E_-)/\sqrt{2}, \; E_y =
(E_+-E_-)/(i\sqrt{2}).
\end{eqnarray}

The solution (\ref{Efield}) written in Cartesian coordinates has the
form

\begin{eqnarray}
\label{REfield}
E_x(x,y,z,t)=\frac{2}{1+\tilde{r}^2}\sin\left[\Delta k_z\left(v_{ph}-v_{gr}\right)t\right]\nonumber\\
\\
E_y(x,y,z,t)=\frac{2}{1+\tilde{r}^2}\cos\left[\Delta
k_z\left(v_{ph}-v_{gr}\right)t\right].\nonumber
\end{eqnarray}
The $3D+1$ Lorentz type solution (\ref{REfield}), presented in
Cartesian coordinates,  gives  oscillation of the electrical vector
$\vec{E}=(E_x,E_y,0)$ in the ($x,y$) plane. It can be seen directly
that the frequency of oscillation is equal to the carrier to
envelope frequency $\omega_{cef}=k_0\left(v_{ph}-v_{gr}\right)$. The
corresponding longitudinal spatial carrier to envelope wave number
is $k_{cef}=\omega_{cef}/v_{gr}$.  Detailed investigation on the
evolution of  the solution (\ref{REfield}) at distances more than
$10-20$ meters from the initial point, reveals weak
self-compression. That is why we consider  the filament in this zone
as closer to a weak Rogue wave. The evolution  of the profiles of
the electrical field components $\Re (E_x)$-Fig. 1a and
$\Re(E_y)$-Fig. 1b. are plotted in Fig. 1. The periodical exchange
of energy between components, due to nonlinear mechanism, leads to
rotation of the  electrical field  vector in a plane orthogonal to
the direction of propagation, with time period $T_{cef}\simeq
1-2\times10^{-10}$ $s$ and spatial period $\Lambda_{cef}\simeq 3-6$
$cm$.

\section{Conclusions}
The starting point of our investigation in this paper is the fact,
that the generalized nonlinear polarization (\ref{NLTH}) arise new
polarization state in circular basis. This polarization state leads
to periodical exchange of energy between the electrical components
$E_x$ and $E_y$ of a laser pulse.  To derive the corresponding
amplitude equations associated with this polarization, we take into
account the orthogonality in the complex plane of the left-hand and
right-hand circular components. We find that the obtained system of
amplitude equations (\ref{SYSCIRK}) has exact (3D+1) Lorentz type
soliton solutions (\ref{soliton1}). Our soliton solution is obtained
for pulses which satisfy the  additional condition $\Delta
k_z\approx k_0$. The diffraction of broad-band pulses is not the
Fresnel one \cite{NAUM, DAKKOV}, which leads to the conclusion that
the soliton appears as a balance between semi-spherical (Fraunhofer
type) diffraction and  nonlinear self-focusing. The solution gives
also a rotation of the vector of the electrical field with the
carrier to envelope frequency.

\end{document}